\input{psfig}
\documentclass[oldversion]{aa}
\input epsf
\input psfig.sty
\newcommand{\be}{\begin{equation}}
\newcommand{\ee}{\end{equation}}
\newcommand{\bea}{\begin{eqnarray}}
\newcommand{\eea}{\end{eqnarray}}
\begin{document}
\title{
Some improvements to the spherical collapse model} 
\author{A. Del Popolo\inst{1,2}
}
\titlerunning{Some improvements to the spherical collapse model}
\authorrunning{A. Del Popolo}
\date{}
\offprints{A. Del Popolo, E-mail:antonino.delpopolo@boun.edu.tr}
\institute{
$^1$ Bo$\breve{g}azi$\c{c}i University, Physics Department,
     80815 Bebek, Istanbul, Turkey\\
$2$ Dipartimento di Matematica, Universit\`{a} Statale di Bergamo,
  via dei Caniana, 2,  24127, Bergamo, ITALY\\
$3$ Istanbul Technical University, Ayazaga Campus,  Faculty of Science and Letters,  34469 Maslak/ISTANBUL, Turkey\\
}


\abstract{
I study the joint effect of dynamical friction, tidal torques and cosmological constant on 
clusters of galaxies formation.
I show that within
high-density environments, such as rich clusters of galaxies,
both dynamical friction and tidal torques slows down the collapse of low-$\nu$ peaks producing
an observable variation in the time of collapse of the perturbation and,
as a consequence, a reduction in the mass bound to the collapsed
perturbation. 
Moreover, the delay of the collapse produces a tendency
for less dense regions to accrete less mass, with respect
to a classical spherical model, inducing a biasing of over-dense
regions toward higher mass. I show how the threshold of collapse is modified if dynamical friction, tidal torques and a non-zero cosmological constant are taken into account
and I use the Extended Press Schecter (EPS) approach to 
calculate the effects on the mass function. 
Then, I compare the numerical mass function given in Reed et al. (2003) with the theoretical mass function obtained in the present paper.
I show that the barrier obtained in the present paper
gives rise to a better description of the mass function evolution with respect to other previous models (Sheth \& Tormen 1999 (hereafter ST), Sheth \& Tormen 2002 (hereafter ST1)).
}


\maketitle

\section{Introduction}

\noindent
Structure formation in universe is generated through the growth and collapse 
of primeval density perturbations originated from quantum fluctuations (Guth \& Pi 1982; 
Hawking 1982; Starobinsky 1982; Bardeen et al. 1986 (hereafter BBKS)) in an inflationary 
phase of early universe. 
Density perturbations evolve towards non-linear regime because of 
gravitational instability, they break away from general expansion and finally they recollapse at a time $T_{c0}$.
The collapse time, $T_{c0}$, depends on the characteristic of the initial fluctuation field and on the environment in which the perturbation 
is embedded. This last features depends on the cosmological scenario. 
In this context, the last years have shown that the most successful 
model is the $\Lambda$CDM model with $\Omega_m=0.3$ and $\Lambda=0.7$.
The statistics of density fluctuations originated in the inflationary era are Gaussian and 
it can be expressed entirely 
in terms of the power spectrum of density fluctuations.
On average the characteristics of the
density field peaks, e.g., their mass distribution, peculiar velocities,
etc., are completely determined by the spectrum through its moments (at
least during the linear and early non-linear phases of the collapse (BBKS)).
Moreover the isotropy condition imposes that all physical
quantities around density peaks is, on average, spherically symmetric.
However, several years ago it was realized that the density field distributions
around the density peaks, which eventually will give birth to galaxies and
clusters, depart from spherical symmetry and from the average density
profile, producing important consequences on collapse dynamics and formation
of protostructures (Hoffman \& Shaham 1985; Ryden 1988a,b; Heavens \& Peacock
1988; Kashlinsky 1986, 1987; Peebles 1990). A fundamental role in this
context is played by the joint action of tidal torques (coupling shells of
matter which are accreted around a density peak and neighboring
protostructures (Ryden 1988)), and by dynamical friction (White 1976;
Kashlinsky, 1986, 1987, Antonuccio \& Colafrancesco 1994 (hereafter AC)).

According to the previrialization conjecture (Peebles \& Groth 1976,
Davis \& Peebles 1977, Peebles 1990), initial asphericities and tidal interactions between neighboring
density fluctuations induce significant non-radial motions which oppose the
collapse. This means that virialized clumps form later, with respect to the
predictions of the linear perturbation theory or the spherical collapse model (hereafter SM),
and that the initial density contrast, needed to obtain a given final
density contrast, must be larger than that for an isolated spherical
fluctuation.
This kind of conclusion was supported by Barrow \& Silk (1981), Szalay \&
Silk (1983), Villumsen \& Davis (1986), Bond \& Myers (1993a,b)
and Lokas et al. (1996). \\
In particular Barrow \& Silk (1981) and Szalay \& Silk (1983) pointed out
that non-radial motions would slow the rate of growth of the density contrast
by lowering the peculiar velocity and suppress collapse once the system
detaches from general expansion. Villumsen \& Davis (1986) gave examples
of the growth of non-radial motions in N-body simulations.
Arguments based on a numerical least-action method lead Peebles (1990)
to the conclusion that irregularities in the mass distribution,
together with external tides, induce
non-radial motions
that slow down the collapse.
Lokas et al. (1996) used N-body simulations and a weakly non-linear
perturbative approach to study previrialization. They concluded that when
the slope of the initial power spectrum is $n>-1$, non-linear tidal
interactions slow down the growth of density fluctuations and the
magnitude of the effect increases when $n$ is increased. \\
Opposite conclusions were obtained by Hoffman (1986a,1989),
Evrard \& Crone (1992), Bartelmann et al. (1993), Bertschinger \& Jain (1994), Monaco (1995) and Kerscher et al. (2001).
In particular Hoffman (1986a,1989), using the quasi-linear (QL) approximation
(Zel'dovich 1970; Zel'dovich \& Novikov 1983) showed that the shear affects
the dynamics of collapsing objects and it leads to infall velocities that are
larger than in the case of non-shearing ones. Bertschinger \& Jain (1994) put
this result in theorem form, according to which spherical perturbations are
the slowest in collapsing. 
Bartelmann et al. (1993) argued that the collapse does not start from a comoving motion of the perturbation, but that the continuity 
equation requires an initial velocity perturbation directly related to the density perturbation. The effect is that collapse proceeds faster than in the case where the initial velocity perturbation is set equal to zero and the collapse timescale is shortened. 
Kerscher et al. (2001) investigated the inhomogeneous, anisotropic collapse of non-isolated dust objects.
Their collapse model, differently from the pure spherical collapse model, which focuses only on
the mass over–density, considers
also the fluctuations in the expansion rate as well as
the effect of shear fields on the collapse. Their 
results indicate that on large mass–--scales the shear fields,
caused by internal and external mass concentrations, accelerate the collapse.
The N-body simulations by Evrard \& Crone (1992)
did not reproduce previrialization effect, but the reason is due to the fact
that they assumed an $n=-1$ spectrum, differently from the $n=0$ one used by
Peebles (1990) that reproduced the effect. If $n<-1$ the peculiar gravitational
acceleration, $g \propto R^{-(n+1)/2}$, diverges at large $R$ and the
gravitational acceleration moves the fluid more or less uniformly,
generating bulk flows rather than shearing motions. Therefore, its collapse
will be similar to that of an isolated spherical clump. If $n>-1$, the
dominant sources of acceleration are local, small-scale inhomogeneities and
tidal effects will tend to generate non-radial motions and resist gravitational
collapse.
In a more recent paper, Audit et al. (1997) have
proposed some analytic prescriptions to compute
the collapse time along the second and the third principal axes of an
ellipsoid,
by means of the 'fuzzy'  threshold approach.
They pointed out that the formation of real virialized clumps must correspond
to the third axis collapse and that the collapse along this axis
is slowed down by the  effect of the shear
rather than be accelerated by it,
in contrast to its effect on the first axis collapse.
They concluded that spherical collapse is the fastest, in disagreement with
Bertschinger \& Jain's theorem.
This result
is in agreement with Peebles (1990).
The quoted controversy was addressed by 
Del Popolo, Ercan \& Xia (2001) who examined the evolution of non-spherical inhomogeneities in a Einstein-de
Sitter universe, by numerically solving the equations of motion for the principal
axes and the density of a dust ellipsoid.
%
%
They showed that for lower values of $\nu$ ($\nu=2$) the
growth rate enhancement of the density contrast induced by 
the shear is counterbalanced by the effect of angular momentum acquisition.
For $\nu>3$ the effect of angular momentum and shear reduces, and the
evolution of perturbations tends to follow the behavior obtained in the SM.
Del Popolo (2002) studied the role of shear fields on the evolution of density perturbations
by using an analytical approximate solution for the equations of
motion of homogeneous ellipsoids embedded in a homogeneous
background. The equations of motion of a homogeneous ellipsoid (Icke 1973;
White \& Silk 1979 (hereafter WS)) were modified in order to take
account of the tidal field, as done in Watanabe (1993) and then were
integrated analytically, similar to what was done in WS. 
The density contrast at turn-around and the
collapse velocity were found to be reduced with respect to that
found by means of the SM. The reduction increases
with increasing strength of the external tidal field and with
increasing initial asymmetry of the ellipsoids. 

The second physical effect with changes cluster collapse is dynamical friction. Former treatments of the dynamical friction effects on the structure of clusters of galaxies, considering only the component generated 
by the galactic population on the motion of galaxies themselves are due to White (1976) and Kashlinsky (1984, 1986, 1987). AC
recalculated the effect of dynamical friction taking into account the effect of substructure, showing that
dynamical friction delays the collapse of low-$\nu $ peaks
inducing a bias of dynamical nature. 
Because of dynamical friction
under-dense regions in clusters (the clusters outskirts) accrete less mass
with respect to that accreted in absence of this dissipative effect and as a
consequence over-dense regions are biased toward higher mass (AC).
Dynamical friction and non-radial motions acts
in a similar fashion: they delay the shell collapse
consequently inducing a dynamical bias.
Whenever efficient, these mechanisms will generate a physical
selection of those peaks in the initial density field that eventually will
give rise to the observed cosmic structures. 
As a consequence of dynamical friction and tidal torques, one expects changes in the threshold of collapse, the mass function and the correlation function.

In this paper, I shall study how structure formation is changed by the joint effect of 
dynamical friction, tidal torques and a non-zero cosmological constant.
The plan of the paper is as follows. In section 2, I summarize the role of dynamical friction in structure formation, in section 3 the role of 
tidal torques. Section 4 shows how the joint effect of dynamical friction and tidal torques affect cluster formation and evolution. Section 5 deals with the modification of 
the threshold of collapse and section 6 with the modifications induced on the mass function. Section 7 is devoted to conclusions.

\section{Dynamical friction and structure formation. } 

In a hierarchical structure formation model, the large scale cosmic
environment can be represented as a collisionless medium made of a hierarchy
of density fluctuations whose mass, $M$, is given by the mass function $%
N(M,z)$, where $z$ is the redshift. In these models matter is concentrated
in lumps, and the lumps into groups and so on.
In such a material system, gravitational
field can be decomposed into an average field, ${\bf F}_0(r)$, generated
from the smoothed out distribution of mass, and a stochastic component, $%
{\bf F}_{stoch}(r)$, generated from the fluctuations in number of the field
particles. 
The stochastic component of the gravitational field is specified assigning a
probability density, $W({\bf F})$, (Chandrasekhar \& von Neumann 1942). In
an infinite homogeneous unclustered system $W({\bf F})$ is given by
Holtsmark distribution (Chandrasekhar \& von Neumann 1942) while in
inhomogeneous and clustered systems $W({\bf F})$ is given by Kandrup (1980)
and Antonuccio-Delogu \& Barandela (1992) respectively. The stochastic
force, ${\bf F}_{stoch}$, in a self-gravitating system modifies the motion
of particles as it is done by a frictional force. In fact a particle moving
faster than its neighbors produces a deflection of their orbits in such a
way that average density is greater in the direction opposite to that of
traveling causing a slowing down in its motion. Following Chandrasekhar
\& von Neumann's (1942) method, the frictional force which is experienced by
a body of mass $M$ (galaxy), moving through a homogeneous and isotropic
distribution of lighter particles of mass $m$ (substructure), having a
velocity distribution $n(v)$ is given by: 
\begin{equation}
M\frac{d{\bf v}}{dt}=-4\pi G^2M^2n(v)\frac{{\bf v}}{v^3}\log \Lambda \rho 
\label{eq:cha}
\end{equation}
where $\log \Lambda $ is the Coulomb logarithm, $\rho $ the density of the
field particles (substructure). \\ A more general formula is that given by
Kandrup(1980) in the hypothesis that there are no correlations among random
force and their derivatives: 
\begin{equation}
{\bf F}=-\eta {\bf v}=-\frac{\int W(F)F^2T(F)d^3F}{2<v^2>}{\bf v}
\end{equation}
where $\eta $ is the coefficient of dynamical friction, $T(F)$ the average
duration of a random force impulse, $<v^2>$ the characteristic speed of a
field particle having a distance $r\simeq (\frac{GM}F)^{1/2}$ from a test
particle (galaxy). This formula is more general than Eq. (\ref{eq:cha})
because the frictional force can be calculated also for inhomogeneous
systems when $W(F)$ is given. If the field particles are distributed
homogeneously the dynamical friction force is given by: 
\begin{equation}
F=-\eta v=-\frac{4.44G^2m_a^2n_a}{[<v^2>]^{3/2}}\log \left\{ 1.12\frac{<v^2>%
}{Gm_an_a^{1/3}}\right\} 
\end{equation}
(Kandrup 1980), where $m_a$ and $n_a$ are respectively the average mass and
density of the field particles. Using virial theorem we also have: 
\begin{equation}
\frac{<v^2>}{Gm_an_a^{1/3}}\simeq \frac{M_{tot}}m\frac
1{n^{1/3}R_{sys}}\simeq N^{2/3}
\end{equation}
where $M_{tot}$ is the total mass of the system, $R_{sys}$ its radius and $N$
is the total number of field particles. The dynamical friction force can be
written as follows: 
\begin{equation}
F=-\eta v=-\frac{4.44[Gm_an_{ac}]^{1/2}}N\log \left\{ 1.12N^{2/3}\right\}
\frac v{a^{3/2}}=-\epsilon_o \frac v{a^{3/2}}
\end{equation}
where $N=\frac{4\pi }3R_{sys}^3n_a$ and $n_{ac}=n_a\times a^3$ is the
comoving number density of peaks of substructure of field particles. This
last equation supposes that the field particles generating the stochastic
field are virialized. This is justified by the previrialization hypothesis
(Davis \& Peebles 1977). \\ To calculate the dynamical evolution of the
galactic component of the cluster it is necessary to calculate the number
and average mass of the field particles generating the stochastic field. \\ %
The protocluster, before the ultimate collapse at $z\simeq 0.02$, is made of
substructure having masses ranging from $10^6-10^9M_{\odot }$ and from
galaxies. I suppose that the stochastic gravitational field is generated
from that portion of substructure having a central height $\nu $ larger than
a critical threshold $\nu _c$. This latter quantity can be calculated
(following AC) using the condition that the peak radius, $r_{pk}(\nu \ge \nu
_c),$ is much less than the average peak separation $n_a(\nu \ge \nu
_c)^{-1/3}$, where $n_a$ is given by the formula of BBKS for the upcrossing
points: 
\begin{eqnarray}
n_{ac}(\nu \ge \nu _c)=\frac{\exp (\nu _c^2/2)}{(2\pi )^2}(\frac \gamma
{R_{*}})^3 [ \nu _c^2-1+ \nonumber \\
\frac{4\sqrt{3}}{5\gamma ^2(1-5\gamma ^2/9)^{1/2}}
\exp ( -5\gamma ^2\nu _c^2/18)] 
\end{eqnarray}
where $\gamma $, $R_{*}$ are parameters related to moments of the power
spectrum (BBKS Eq. ~4.6A). The condition $r_{pk}(\nu \ge \nu _c)<0.1n_a(\nu
\ge \nu _c)^{-1/3}$ ensures that the peaks of substructure are point like.
Using the radius for a peak: 
\begin{equation}
r_{pk}=\sqrt{2}R_{*}\left[ \frac 1{(1+\nu \sigma _0)(\gamma ^3+(0.9/\nu
))^{3/2}}\right] ^{1/3}
\end{equation}
(AC), I obtain a value of $\nu _c=1.3$ and then we have $n_a(\nu \ge \nu
_c)=50.7Mpc^{-3}$ 
($\gamma =0.4$, $R_{*}=50Kpc$) and $m_a$ is given by: 
\begin{equation}
m_a=\frac 1{n_a(\nu \ge \nu _c)}\int_{\nu _c}^\infty m_{pk}(\nu )N_{pk}(\nu
)d\nu =10^9M_{\odot }
\end{equation}
(in accordance with the result of AC), where $m_{pk}$ is given in Peacock $\&
$ Heavens (1990) and $N_{pk}$ is the average number density of peak (BBKS
Eq. ~4.4).
Galaxies and Clusters of galaxies are correlated systems whose autocorrelation function, $\xi(r)$,
can be expressed, 
in a power law form (Peebles 1980; Bahcal \& Soneira 1983; Postman et al. 1986; Davis \& Peebles 1983; Gonzalez et al. 2002).
%
%
The description of dynamical friction in these systems need to
use a distribution of the stochastic forces, $W(F)$, taking account of
correlations. In this last case the coefficient of dynamical friction,
$ \eta$, may be calculated using the equation:
\begin{equation}
\eta=\int  d^{3} {\bf F} W(F) F^{2} T(F)/(2<v^{2}>)
\end{equation}
and using Antonuccio \& Atrio (1992) distribution:
\begin{equation}
W(F)=\frac{1}{2 \pi^2 F} \int_{0}^{\infty} dk k sin(kF)A_{f}(k)
\end{equation}
where $A_{f}$, which is a linear integral function of the correlation function $ \xi(r)$,
is given in the quoted paper (Eq. 36). 

\section{Tidal torques and structure formation.}

The explanation of galaxies spins gain through tidal torques was pioneered
by Hoyle (1949). Peebles (1969) performed the first detailed calculation of the
acquisition of angular momentum in the early stages of protogalactic
evolution. More recent analytic computations (White 1984, Hoffman 1986,
Ryden 1988a; Eisenstein \& Loeb 1995; Catelan \& Theuns 1996a, b) and numerical simulations (Barnes \& Efstathiou 1987) have
re-investigated the role of tidal torques in originating galaxies angular
momentum. 
One way to study the variation of angular momentum with radius in
a galaxy is that followed by Ryden (1988a). In this approach the protogalaxy
is divided into a series of mass shells and the torque on each mass shell is
computed separately. The density profile of each proto-structure is
approximated by the superposition of a spherical profile, $\delta (r)$, and
a random CDM distribution, ${\bf \varepsilon (r)}$, which provides the
quadrupole moment of the protogalaxy. 
As shown by Ryden (1988a) the net rms torque on a
mass shell centered on the origin of internal radius $r$ and thickness $%
\delta r$ is given by: 
\begin{eqnarray}
\langle |\tau |^2\rangle ^{1/2}=\sqrt{30}\left( \frac{4\pi }5G\right) 
[\langle a_{2m}(r)^2\rangle \langle q_{2m}(r)^2\rangle \nonumber \\
-\langle a_{2m}(r)q_{2m}^{*}(r)\rangle ^2] ^{1/2}  \label{eq:tau}
\end{eqnarray}
where $q_{lm}$, the multipole moments of the shell and $a_{lm}$, the tidal
moments, are given by: 
\begin{equation}
\langle q_{2m}(r)^2\rangle =\frac{r^4}{\left( 2\pi \right) ^3}M_{sh}^2\int
k^2dkP\left( k\right) j_2\left( kr\right) ^2
\end{equation}
\begin{equation}
\langle a_{2m}(r)^2\rangle =\frac{2\rho _b^2r^{-2}}\pi \int dkP\left(
k\right) j_1\left( kr\right) ^2
\end{equation}
\begin{equation}
\langle a_{2m}(r)q_{2m}^{*}(r)\rangle =\frac r{2\pi ^2}\rho _bM_{sh}\int
kdkP\left( k\right) j_1\left( kr\right) j_2(kr)
\end{equation}
where $M_{sh}$ is the mass of the shell, $j_1(r)$ and $j_2(r)$ are the
spherical Bessel function of first and second order while the power spectrum 
$P(k)$ is given by, BBKS (equation~(G3)): 
\begin{eqnarray}
T(k) &=& \frac{[\ln \left( 1+2.34 q\right)]}{2.34 q}
\nonumber \\
& & 
\cdot [1+3.89q+
(16.1 q)^2+(5.46 q)^3+(6.71)^4]^{\frac{-1}{4}}
%
%
\label{eq:ma5}
\end{eqnarray}
(where 
$q=\frac{k\theta^{1/2}}{\Omega_{\rm X} h^2 {\rm Mpc^{-1}}}$.
Here $\theta=\rho_{\rm er}/(1.686 \rho_{\rm \gamma})$
represents the ratio of the energy density in relativistic particles to
that in photons ($\theta=1$ corresponds to photons and three flavors of
relativistic neutrinos).
The power spectrum was normalized to reproduce the observed abundance of rich 
cluster of galaxies (e.g., Bahcal \& Fan 1998).
Filtering the spectrum on cluster scales, $R_f=3h^{-1}Mpc$, I have obtained
the rms torque, $\tau (r)$, on a mass shell using Eq. (\ref{eq:tau}) then I
obtained the total specific angular momentum, $h(r,\nu )$, acquired during
expansion integrating the torque over time (Ryden 1988a Eq. 35): 
\begin{eqnarray}
h(r,\nu )=\frac 13\left( \frac 34\right) ^{2/3} \nonumber \\
\frac{\tau _ot_0}{M_{sh}}%
\overline{\delta }_o^{-5/2}\int_0^\pi \frac{\left( 1-\cos \theta \right) ^3}{%
\left( \vartheta -\sin \vartheta \right) ^{4/3}}\frac{f_2(\vartheta )}{%
f_1(\vartheta )-f_2(\vartheta )\frac{\delta _o}{\overline{\delta _o}}}%
d\vartheta   \label{eq:ang}
\end{eqnarray}
the functions $f_1(\vartheta )$, $f_2(\vartheta )$ are given by Ryden
(1988a) (Eq. 31) while the mean over-density inside the shell, $\overline{%
\delta }(r)$, is given by Ryden (1988a): 
\begin{equation}
\overline{\delta }(r,\nu )=\frac 3{r^3}\int_0^\infty d\sigma \sigma ^2\delta
(\sigma )
\label{eq:denss}
\end{equation}
where $\delta(r)=\frac{\rho(r)-\rho_b}{\rho_b}$.
%
%
As showed by Ryden (1988a), the rms specific angular momentum, 
$h(r,\nu )$, increases with distance $r$ while peaks of greater $\nu $
acquire less angular momentum via tidal torques. This is the angular
momentum-density anticorrelation showed by Hoffman (1986). This effect
arises because the angular momentum is proportional to the gain at turn
around time, $t_m$, which in turn is proportional to $\overline{\delta }%
(r,\nu )^{-\frac 32}\propto \nu ^{-3/2}$.

\section{Modification of collapse.}

Tidal torques and dynamical friction acts in a similar fashion. As reported in 
the introduction AC calculated the effect of dynamical friction taking into account the effect of substructure, showing that
dynamical friction delays the collapse of low-$\nu $ peaks
inducing a bias of dynamical nature. Similarly non-radial motions would slow the rate of growth of the density contrast
by lowering the peculiar velocity and suppress collapse once the system
detaches from general expansion. 
In fact, in the central regions of a density peak ($r\leq 0.5R_f$) the velocity
dispersion attain nearly the same value 
while at larger radii ($r\geq R_f$) the radial component is lower than the
tangential component. This means that motions in the outer regions are
predominantly non-radial and in these regions the fate of the infalling
material could be influenced by the amount of tangential velocity relative
to the radial one. This can be shown writing the equation of motion of a
spherically symmetric mass distribution with density $n(r)$ (Peebles 1993): 
\begin{equation}
\frac \partial {\partial t}n\langle v_r\rangle +\frac \partial {\partial
r}n\langle v_r^2\rangle +\left( 2\langle v_r^2\rangle -\langle v_\vartheta
^2\rangle \right) \frac nr+n(r)\frac \partial {\partial t}\langle v_r\rangle
=0  \label{eq:peeb}
\end{equation}
where $\langle v_r\rangle $ and $\langle v_\vartheta \rangle $ are,
respectively, the mean radial and tangential streaming velocity. Eq. (\ref
{eq:peeb}) shows that high tangential velocity dispersion $(\langle
v_\vartheta ^2\rangle \geq 2\langle v_r^2\rangle )$ may alter the infall
pattern. The expected delay in the collapse of a perturbation, due to
non-radial motions, dynamical friction and also taking account of a non-zero cosmological constant, may be
calculated solving the equation for the radial acceleration (Kashlinsky
1986, 1987; AC; Peebles 1993): 
\begin{equation}
\frac{dv_r}{dt}=\frac{L^2(r,\nu )}{M^2r^3}-g(r) -\eta \frac{dr}{dt}+ \frac{\Lambda}{3}r \label{eq:coll}
\end{equation}
where $L(r,\nu )$ is the angular momentum, $g(r)$ the acceleration, and $\Lambda$ the cosmological constant.
\begin{figure}
\psfig{file=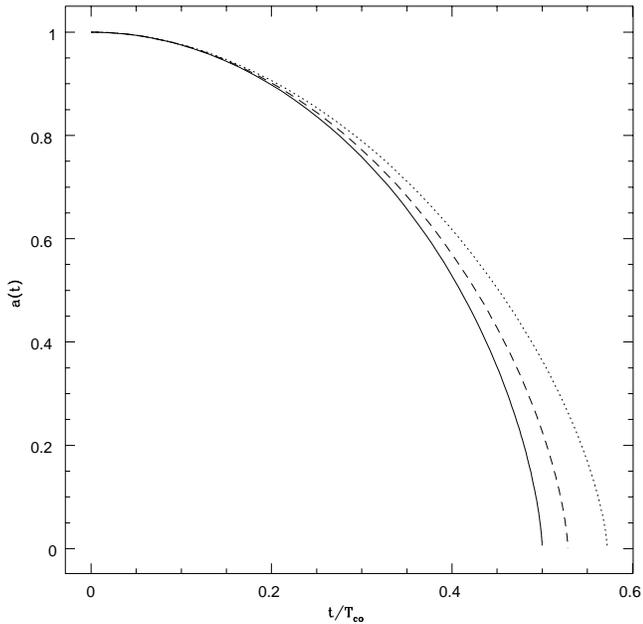,width=9cm,height=9cm}
\caption[]{The time evolution of the expansion parameter. The solid 
line is a(t) for the SM; the dashed line is a(t) taking account only
dynamical friction; the dotted line is a(t) taking account of the cumulative
effect of non-radial motions and dynamical friction in the case of a $\nu=2$ peak.}
\end{figure}
Writing the proper radius of a shell in terms of the expansion parameter, $%
a(r_i,t)$: 
\begin{equation}
r(r_i,t)=r_ia(r_i,t)
\end{equation}
remembering that 
\begin{equation}
M=\frac{4\pi }3\rho _b(r_i,t)a^3(r_i,t)r_i^3
\end{equation}
and that $\rho _b=\frac{3H_0^2}{8\pi G}$, where $H_0$ is the Hubble constant
and assuming that no shell crossing occurs so that the total mass inside
each shell remains constant, that is:
\begin{equation}
\rho (r_i,t)=\frac{\rho _i(r_i,t)}{a^3(r_i,t)}
\end{equation}
the Eq. (\ref{eq:coll}) may be written as: 
\begin{equation}
\frac{d^2a}{dt^2}=-\frac{H^2(1+\overline{\delta })}{2a^2}+\frac{4G^2L^2}{%
H^4(1+\overline{\delta })^2r_i^{10}a^3} -\eta \frac{da}{dt}+ \frac{\Lambda}{3}a 
\label{eq:sec}
\end{equation}
\begin{figure}
\centerline{\hbox{
2a
\psfig{file=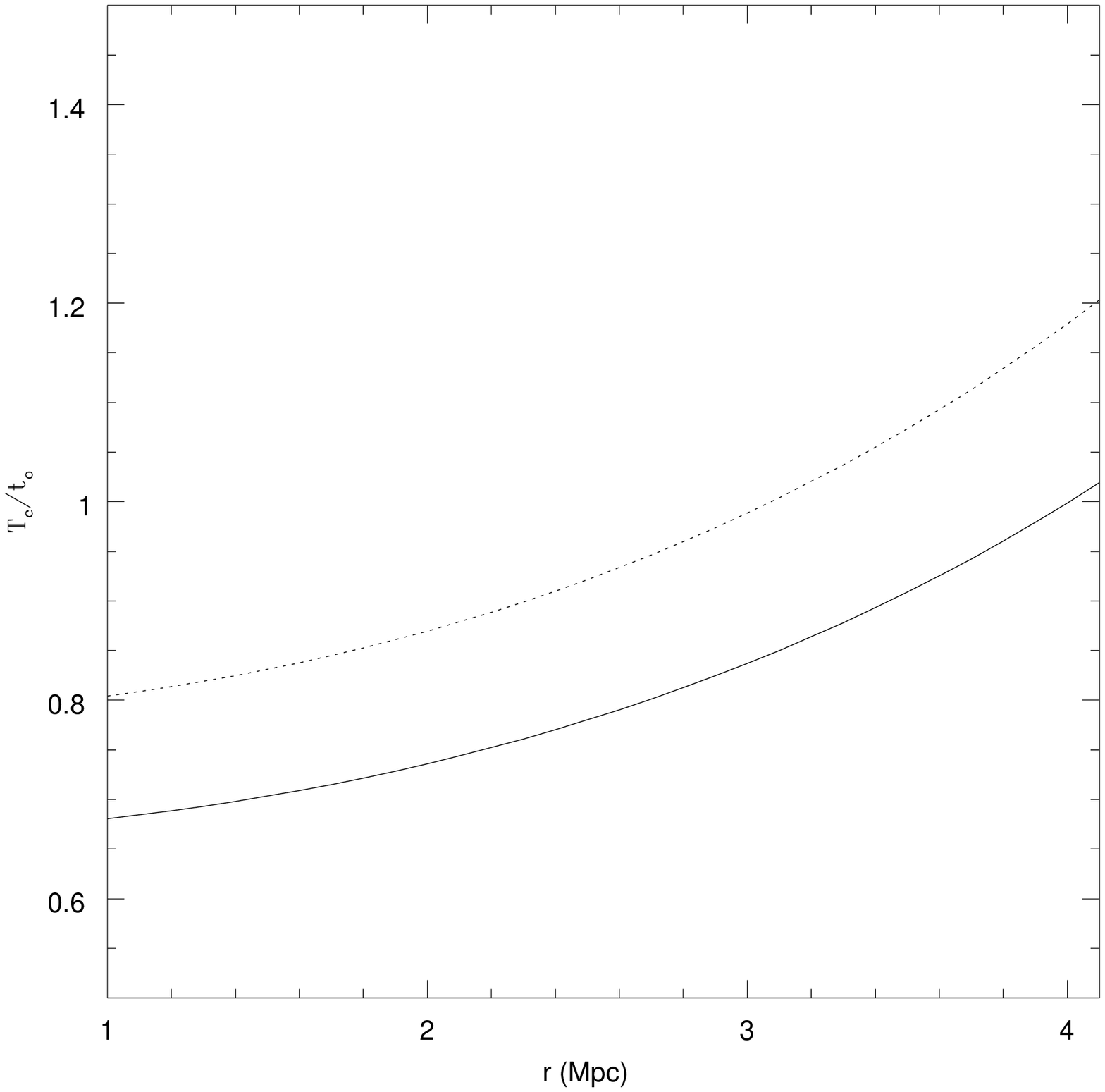,width=9cm}  
}}
\centerline{\hbox{
2b
\psfig{file=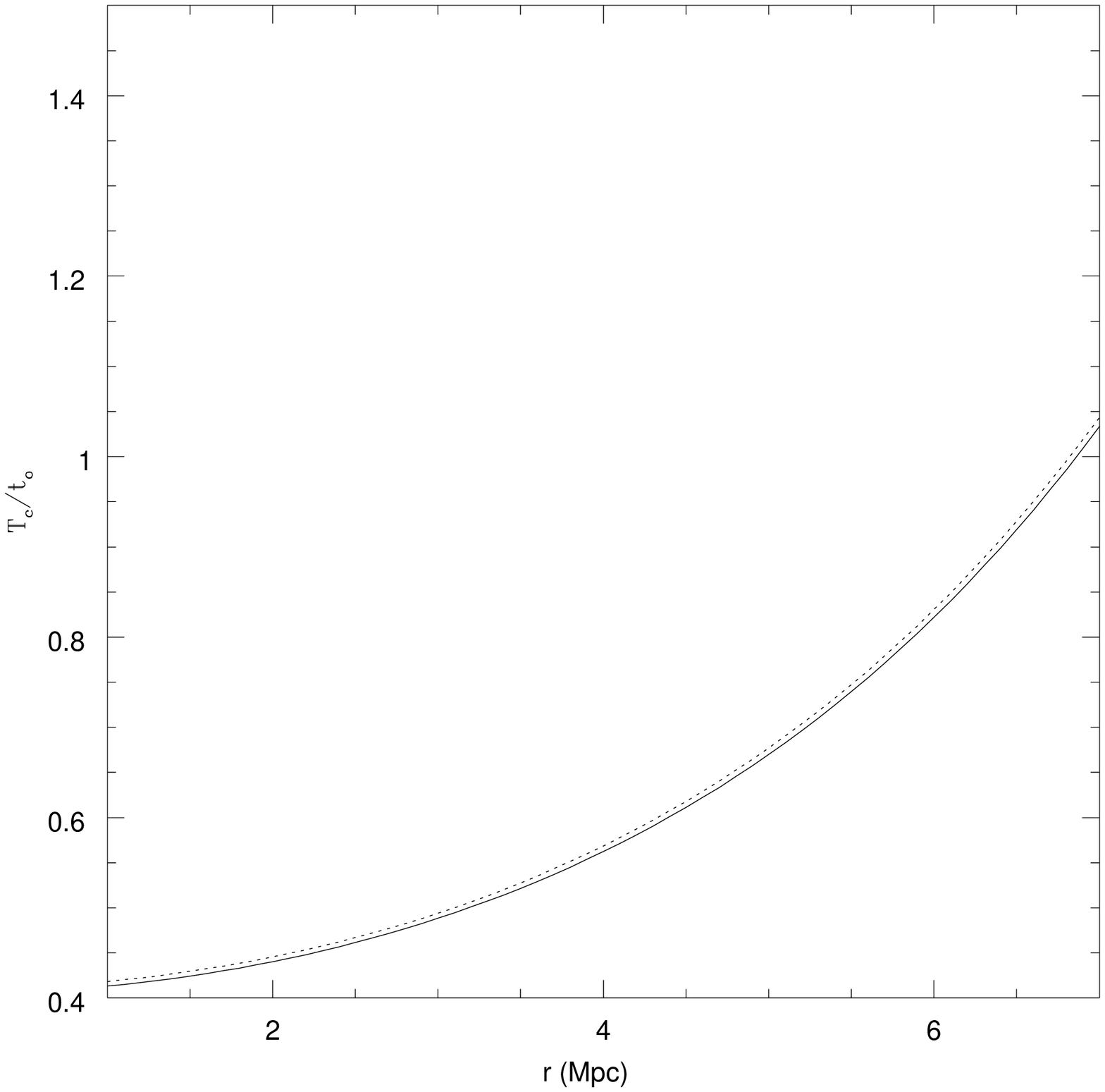,width=9cm} 
}}
\caption[]{ Fig. 2a. The time of collapse of a shell of matter in units of the age of the
universe $t_{o}$ for $\nu=2$ (dotted line) compared with SM
(solid line).
Fig. 2b. The time of collapse of a shell of matter in units of the age of the
universe $t_{o}$ for $\nu=4$ (dotted line) compared with SM (solid line).
}
\end{figure}
Eq. (\ref{eq:sec}) does not take into account back-reaction terms, which have been taken into account by Kerscher et al. (2001).
On large mass–--scales this generic model leads to changes in the pure top-hat model. One consequence of the Kerscher et al. (2001) model is the prediction of more collapsed objects at high-mass
scales, which, as we shall see in the next sections,  
leads to conclusions similar to the model of the present paper.

The equation (\ref{eq:sec}) 
may be
solved using the initial conditions: $(\frac{da}{dt})=0$, $a=a_{max}\simeq 1/%
\overline{\delta }$ and using the function $h(r,\nu )=L(r,\nu )/M_{sh}$
found in Section 2 to obtain a(t) and the time of collapse, $T_c(r,\nu )$. As shown by Gunn \& Gott (1972), this last quantity in the case of a pure 
SM (namely when tidal torques, dynamical friction and cosmological constant are not taken into account)
is given by:
\begin{equation}
T_{c0}(r,\nu )=\frac{\pi}{H_i [\overline{\delta }(r,\nu )]^{3/2}}
\footnote{As we told in the introduction, the inclusion of the peculiar velocity field changes the collapse as:
$H_i T_{c0} \simeq \frac{\pi}{(c \delta_i -\epsilon_i)^{3/2}}$
where $c$ and $\epsilon_i$ are defined in Bartelmann et al. (1993). For $\epsilon_i=0$, the collapse is shortened by a factor of $(3/5)^{3/2}$.
}
\label{eq:beef}
\end{equation} 
In Fig. 1, I show the effects of non-radial motions and dynamical friction
separately, in the case of a $\nu=2$ peak. As displayed non-radial motions have a larger effect on the
collapse delay with respect to dynamical friction.
In Fig. 2, I compare the results for the time of collapse, $T_c$, for $\nu
=2$, 4 with the time of collapse of the classical SM (Eq. \ref{eq:beef}).
As shown the presence of non-radial motions produces an increase in the time
of collapse of a spherical shell. The collapse delay is larger for low value
of $\nu $ and becomes negligible for $\nu \geq 3$. This result is in
agreement with the angular momentum-density anticorrelation effect: density
peaks having low value of $\nu $ acquire a larger angular momentum than high 
$\nu $ peaks and consequently the collapse is more delayed with respect to
high $\nu $ peaks. 

Given $T_c(r,\nu )$, I also calculated the total mass
gravitationally bound to the final non-linear configuration. 
There are at least two criteria to establish the bound region to a perturbation $\delta
(r)$: a statistical one (Ryden 1988b), and a dynamical one (Hoffman \&
Shaham 1985), summarized in the following. 

In biased galaxy formation theory structures form around the 
local maxima of the density field. Every density peak binds 
a mass $M$ that can be calculated when we know the binding radius 
of the density peak. The radius of the bound region 
for a chosen density profile $\overline{\delta}( r)$ may be 
obtained in several ways. A first criterion is statistic. The binding 
radius of the region, $ r_{b}$, is given by the solution of the equation:  
\begin{equation}
< \overline{\delta} (r)> = 
< ( \overline{\delta} -<\overline{\delta}>)^{2}>^{1/2}
\end{equation}
(Ryden 1988). At radius $ r << r_{b}$ the motion of particles is 
predominant toward the peak while when $r >> r_{b} $ the particle 
is not bound to the peak. Another criterion 
that can be used is dynamical. It supposes that the binding radius 
is given by the condition that a shell collapse in a time, $ T_{c}$, 
smaller than the age of the universe $ t_{0}$: 
\begin{equation}
T_{c} (r) \leq t_{0} \label{eq:temp}
\end{equation} 
(Hoffmann \& Shaham 1985). 
This last criterion, differently from 
the previous one, contains some prescriptions particularly connected with the 
physics of the collapse process of a shell. For this reason 
I used it to calculate the binding radius. The time of collapse, $ T_{c}(r)$, 
at radius $ r$ can be obtained solving numerically Eq. (\ref{eq:sec})
for different values of $ \overline{\delta_{i}}$, the initial overdensity, 
from a given density profile $ \overline{\delta}(r)$. I use the average density 
profile given by BBKS: 
\begin{equation}
\delta(r)= A \left\{ \frac{\nu \xi(r)}{ \xi(0)^{1/2}}- 
\frac{\theta( \nu \gamma, \gamma)}{\gamma \xi(0)^{1/2}(1-\gamma^{2})}
\left[\gamma^{2} \xi(r) +\frac{ R_{\ast}^{2} 
\bigtriangledown^{2} \xi(r)}{3}\right]\right\} \label{eq:dens}
\end{equation} 
where A is a constant given by the normalization of the 
perturbation spectrum, $ P(k)$,  $ \nu = \frac{\delta_{\rm c0}}{\sigma(M)} $, where $\delta_{\rm co}=1.686$ is the critical threshold for a SM, 
$\sigma(M)$ is the r.m.s. density fluctuation on the mass scale 
$M$, $ \xi(r)$ is the correlation function of two points, $ \gamma$ and $ R_{\ast}$ 
two constants obtainable from the spectrum (see BBKS) and 
finally $ \theta ( \gamma \nu, \gamma)$ is a function given in 
the quoted paper (eq. 6.14). Given the average density 
profile the average density inside the radius $ r$ in  a spherical 
perturbation is given by Eq. (\ref{eq:denss}). 

Finally, I calculated the binding radius, $r_{b}(\nu )$, for a SM, calculating $T_{\rm co}(r)$ by means of   
Eq.~(\ref{eq:beef}) 
and the density profile given in Eq.~(\ref{eq:dens}) and then applying the condition $T_{\rm co}(r) \leq t_o$.
I repeated the calculation for 
$ 1.7 <\nu< 4 $.
Then I repeated the calculation using $T_c(r)$, the collapse time that takes into account non-radial motions and dynamical friction.
I found a relation between $\nu $ and the mass of the cluster using the
equation: $M=\frac{4\pi }3r_b^3\rho _b$.
The result is the plot in Fig. ~3 for the binding radius $ r_{b} $ versus $ \nu$.

In fig. 3, I compare the peak
mass obtained from SM, using Hoffman \& Shaham's (1985) criterion,
with that obtained from the model taking into account non-radial motions, dynamical friction and $\Lambda \neq 0$. As
shown for high values of $\nu $ ($\nu \geq 3$) the two models give the same
result for the mass while for $\nu <3$ the effect of non-radial motions
produces less bound mass with respect to SM. 
Decreasing the effect of non-radial
motions produces a decrease in the bound mass.

The situation represented in the previous three figures may be summarized as follows: dynamical friction and non-radial motions delays the collapse of 
perturbations. Both effects act in the direction of delaying structure collapse, so that their effects add.
The effects have a similar magnitude, but non-radial motions
induce a slight larger delay in collapse. As a consequence of this delay of collapse the matter bound to structures is less than what expected in the case of SM.

\section{The threshold of collapse $\delta_{\rm c}$.
}

In this section, I am going to show how dynamical friction and tidal fields influence 
the critical overdensity threshold for the
collapse, $\delta _c$, which is not constant as in a SM
but it depends on mass.
An analytic determination of $\delta _c(\nu )$ can be obtained following a
technique similar to that used by Bartlett \& Silk (1993). 

\begin{figure}[ht]
\psfig{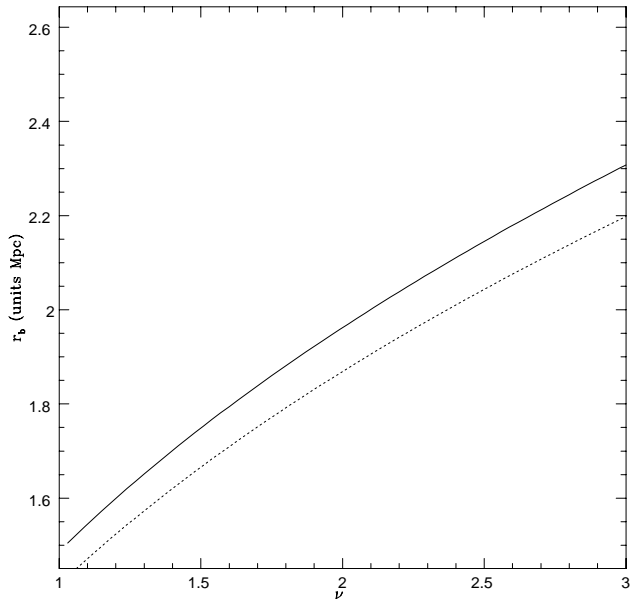}
\caption[h]{Variation of the binding radius $ r_{b} $ with $ \nu $.
The solid line is the binding radius in the SM,
while the dashed line is the same as in presence of non-radial motions dynamical friction and $\Lambda \neq 0$, for $\nu=2$.
}
\end{figure}

Using Eq.~(\ref{eq:sec}) it is possible to obtain the value of the expansion parameter of
the turn around epoch, $a_{max}$, which is characterized by the condition $\frac{da}{dt}=0$. Using the relation between $v$ and $\delta _i$, in linear
theory (Peebles 1980), I find:

\begin{eqnarray}
B(M)&=&\delta _{\rm c}=\delta _{\rm co}
\nonumber \\
& &
\left[ 1+
\int_{r_{\rm i}}^{r_{\rm ta}}  \frac{r_{\rm ta} L^2 \cdot {\rm d}r}{G M^3 r^3}+\frac{\lambda_{o}}{1-\mu(\delta)}+\Lambda \frac{r_{\rm ta} r^2}{6 G M}
\right] 
\label{eq:barriera} 
\end{eqnarray}
%
%
where $\delta _{co}=1.686$ is the critical threshold for SM, $r_{\rm i}$ is the initial radius, $r_{\rm ta}$ is the turn-around radius, 
$\lambda_{o}=\epsilon_o T_{co}$ and $ \mu(\delta)$ is given in Colafrancesco, Antonuccio \& Del
Popolo (1995) (Eq. 29). The quantity $L$ appearing in Eq. ~(\ref{eq:barriera}) is 
the total angular momentum acquired by the proto-structure during evolution. In order to calculate $L$, it is possible to use the same model
described in Sec. 3.
(more hints on the model and some of the model limits can be found in Del Popolo, Ercan \& Xia 2001).
%
%
%
%
The result of the calculation is shown
in Fig. 5, where I plot $\delta_{\rm c}(\nu)$ obtained by
means of the model of the present paper together with that obtained by ST using
an ellipsoidal collapse model. The dashed line
represents $\delta_{\rm c}(\nu)$ obtained with the present model, while the
solid line that of ST. Both models show that the threshold
for collapse
decreases with mass and when $\nu $ $>3$ the
threshold assume the typical value of the SM.
In other words, this means that, in order to form
structure, more massive peaks must
cross a lower threshold,
$\delta_c(\nu)$, with respect to under-dense ones.
At the same time, since the
probability to find high peaks is larger in more dense regions, 
this means that, statistically, in order to form structure, 
peaks in more dense
regions may have a lower value of the threshold, $\delta_c(\nu)$, with respect
to those of under-dense regions.
This is due to
the fact that less massive objects are more influenced by external tides, and
consequently they must be more overdense to collapse by a given time.
In fact,
the angular momentum acquired by a shell centered on a peak
in the CDM density distribution is anti-correlated with density: high-density
peaks acquire less angular momentum than low-density peaks
(Hoffman 1986; Ryden 1988).
A larger amount of
angular momentum acquired by low-density peaks
(with respect to the high-density ones)
implies that these peaks can more easily resist
gravitational collapse and consequently it is more difficult for them to form
structure.
This is in agreement with  
Audit et al. (1997), Peebles (1990), which pointed out that the
gravitational collapse 
is slowed down by the  effect of the shear
rather than fastened by it (as sustained by other authors).
Therefore, on small scales, where the shear is statistically greater,
structures need, on average, a higher  density contrast to collapse. This results in a tendency for less dense
regions to accrete less mass, with respect to a classical SM,
inducing a {\it biasing} of over-dense regions towards higher mass.

\begin{figure}
\psfig{file=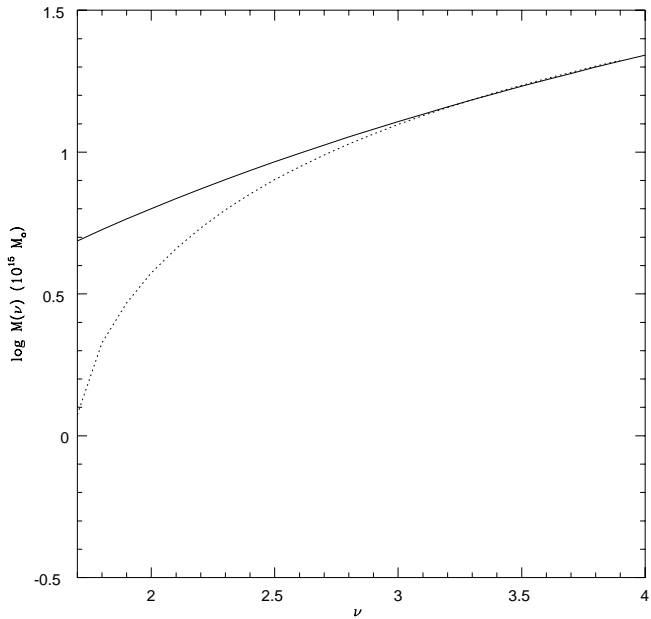,width=9cm,height=9cm}
\caption[]{The mass accreted by a collapsed perturbation, in units of
$10^{15}M_{\odot}$, taking into account non-radial motions, dynamical friction and a non-zero cosmological constant (dotted line)  
compared to SM mass (solid line).}
\end{figure}

\begin{figure}[ht]
\psfig{file=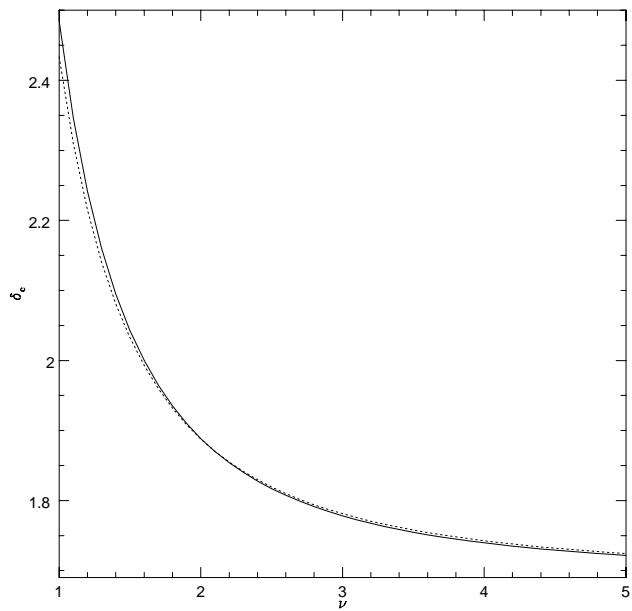,width=9cm,height=9cm}
\caption[]{The critical threshold, $\delta_{c}(\nu)$ versus $\nu$. The dashed line is obtained with the model of the present paper while the solid line is that of ST.} 
\end{figure}

\section{Mass function}

In cosmology, it is possible to model the statistics of halo formation and mergers, 
in a theory sometimes referred to as extended-Press-Schechter, or
excursion set (see PS, Bond et al. 1991, Lacey \& Cole 1993) by means of a random walk.
As well known, a random walk is a stochastic process consisting of a sequence of uncorrelated discrete steps. 
The idea can be summarized as follows. Suppose that density perturbations are represented by a density field and 
imagine smoothing the density field around this point with a filter, say a top-hat filter,
of a successively smaller radius. In a hierarchical universe such as our own, 
one expects that a plot of the smoothed density versus the size of the filter traces out a random walk.
The overall trend is for the smoothed density to rise with a smaller filter radius, but with significant fluctuations.
 
The excursion set theory postulates a flat threshold or barrier (usually motivated by spherical collapse):
once the random walk crosses the barrier, a halo is declared to have formed with a size given by the
crossing radius. Often the filter radius $R$ is denoted by alternative equivalent quantities, e.g.:
the mass variance $S$. The latter is particularly useful for Gaussian random initial conditions -- note that 
$S$ increases as the radius or mass is decreased. 
A natural question to ask is: given Gaussian random initial conditions, what is the probability that
a random walk first crosses the barrier between $S$ and $S + dS$, denoted by
$f(S) dS$? The quantity $f(S)$ is known as
the first-crossing distribution. In excursion set theory, the first-crossing distribution is directly
related to the halo mass function, a key quantity of interest.

Some years ago, it was found that it useful to consider random walks with a non-flat barrier, which is usually referred to as moving barrier, 
in the sense that the barrier 'moves' with $S$. 
For instance,
Sheth et al. (2001, SMT01 hereafter) found that an ellipsoidal
collapse model suggests a moving barrier for defining
halos, which produces a first-crossing distribution, or halo
mass function, that better matches N-body simulations.
SMT01 estimated the mass function assuming
an anisotropic ellipsoidal collapse, where the time evolution
of the half–--axes is determined by the "Zel'dovich
approximation". Kerscher et al. (2001) generalized
this ansatz to domains not restricted
to be ellipsoidal throughout the evolution.
Several other approaches are also based on the "Zel'dovich approximation",
sometimes the third–--order Lagrangian perturbation
approximation is employed. In these "local" approaches
the collapse of a domain is associated with a diverging local matter density (see e.g. Bartelmann et al. 1993,Monaco
1997a,b, Lee \& Shandarin 1998.)
All these models, similarly to the results obtained in the present paper, for the mass function, agrees on predicting a larger abundance of 
collapsed objects at high-mass scales, similarly to N-body simulations like those of
Jenkins et al. 2001 (J01); R03; Yahagi, Nagashima \& Yoshii (2004) (YNY); Warren  et al. 2005 (W05)). 
ST1 connected the form of the barrier with the form of the
mass function. As shown by ST1, for a given barrier shape, $B(S)$, 
the first crossing distribution is well approximated by:
\begin{equation}
f(S)dS=|T(S)|\exp (-\frac{B(S)^{2}}{2S})\frac{dS/S}{\sqrt{2\pi S}}
\label{eq:distrib}
\end{equation}   
where $T(S)$ is the sum of the first few terms in the Taylor expansion of $B(S)$:
\begin{equation}
T(S)=\sum_{n=0}^{5}\frac{(-S)^{n}}{n!}\frac{\partial ^{n}B(S)}{\partial S^{n}}
\label{eq:expans}
\end{equation}
In the case of the barrier shape given in Eq. (\ref{eq:barriera}) 
the Eqs. ~(\ref{eq:distrib}),(\ref{eq:expans}), give, after truncating the expansion at $n=5$ (see ST),
the multiplicity function, $\nu f(\nu)$: 
\begin{eqnarray}
\nu f(\nu )&=& A _1 \left( 1+\frac{\beta_1 g(\alpha_1)}{\left( a\nu \right) ^{\alpha_1}}
+\frac{\beta_2 g(\alpha_2)}{\left( a\nu \right) ^{\alpha_2}}+\frac{\beta_3 g(\alpha_3)}{\left( a\nu \right) ^{\alpha_3}}
\right) \sqrt{\frac{a\nu }{2\pi }}
\nonumber \\
& &
\exp{\{\frac{-a \nu}{2} \left[ 1+\frac{\beta_1}{\left( a\nu \right) ^{\alpha_1}}
+\frac{\beta_2}{\left( a\nu \right) ^{\alpha_2}}+\frac{\beta_3}{\left( a\nu \right) ^{\alpha_3}}
\right] ^{2}\}}
\label{eq:mia}
\end{eqnarray}
where
\begin{equation}
g(\alpha_i)=
\mid 1-\alpha_i +\frac{\alpha_i (\alpha_i
-1)}{2!}-...-\frac{\alpha_i(\alpha_i-1)\cdot \cdot \cdot
(\alpha_i-4)}{5!} \mid
\end{equation}
where $i=1$ or 2, $\alpha_1=0.585$, $\beta_1=0.46$, $\alpha_2=0.5$ and $\beta_2=0.35$, $\alpha_3=0.4$ and $\beta_3=0.02$, and $a=0.707$.

The ``multiplicity function" is correlated with the usual, more straightforwardly used, ``mass function" as follows.  
Following ST notation, if $f(M,\delta) dM$ denotes the fraction of mass that is contained in collapsed haloes that have mass in the range $M$-$M+dM$, at redshift $z$, and $\delta(z)$ the redshift dependent overdensity, 
the associated ``unconditional" mass function is:
\begin{equation}
n(M,\delta)dM=\frac{{\rho_b}}{M} f(M,\delta) dM
\label{eq:mfu}
\end{equation}

In the excursion set approach, the universal or ``unconditional" mass function, $n(M,z)$, representing 
the average comoving number density of haloes of mass $M$ 
is given by:
\begin{equation}
n(M,z)=\frac{{\rho_b(z)}}{M^{2}}\frac{d\log{\nu }}{d\log M}\nu f(\nu )
\label{eq:universall}
\end{equation}
(Bond et al. 1991).
The function $\nu f(\nu)$ is the ``multiplicity function"
which is obtained by computing the distribution of first crossings, $f(\nu) d \nu$, of a barrier $B(\nu)$, by independent, uncorrelated Brownian motion random walks. 
Multiplicity function and mass function are related by Eq. (\ref{eq:universall}). It is to be noted that in literature sometime the terms mass function and multiplicity function are used as synonymous (e.g. ST, Lin et al. 2002).

With a similar calculation, ST1 found, that  
\begin{equation}
\nu f(\nu)\simeq A_2 \left( 1+\frac{0.094}{\left( a\nu \right) ^{0.6}}\right) \sqrt{\frac{a\nu }{2\pi }}\exp{\{-a\nu \left[ 1+\frac{0.5}{\left( a\nu \right) ^{0.6}}\right] ^{2}/2\}}
\label{eq:sstt1}
\end{equation}
with $A_2 \simeq 1$.
This last result is in good agreement with the fit of the simulated first crossing distribution (ST):
\begin{equation}
\nu f(\nu )d\nu =A_3 \left( 1+\frac{1}{\left( a\nu \right) ^{p}}\right) \sqrt{\frac{a\nu }{2\pi }}\exp (-a\nu /2)
\label{eq:ssttt}
\end{equation}
where $p=0.3$, and $a=0.707$. 

The normalization factor $A_3$ has to satisfy the constraint:
\begin{equation}
\int_0^{\infty} f(\nu) d \nu=1
\end{equation}
and as a consequence it is not an independent parameter, but is expressed in the form:
\begin{equation}
A_3=\left[1+2^{-p} \pi^{-1/2} \Gamma(1/2-p)\right]^{-1}=0.3222
\end{equation}

Recently in order to investigate the functional form of the universal
multiplicity function, several authors (J01; R03; YNY; W05) performed runs of N-body simulations
with high mass resolution and compared them with different multiplicity function.
The result of the previous studies is that even the ST multiplicity function does not give correct prediction in some ranges of mass. 
For example, R03 used a high resolution $\Lambda$CDM numerical simulation to calculate the mass function
of dark matter haloes, and its evolution, down to the scale of dwarf galaxies, back to a redshift of
fifteen, in a 50 $h^{-1}$Mpc volume containing 80 million particles.  
Their low redshift results allow us to probe low $\sigma$ density fluctuations
significantly beyond the range of previous cosmological simulations.
They showed that the ST function provides a good fit to their data, except at very high redshifts, where it significantly overpredicts the halo abundance. At all
redshifts up to $z=10$, the difference is $\leq 10 \%$ for each of our well sampled mass bins.  However, the ST  function begins to
overpredict the number of haloes increasingly with redshift for z$\geq$10, up to $\simeq 50$\% by z$=$15.  The simulation mass
functions appear to be generally steeper than the ST function, especially at high redshifts. 
The ST function fits the simulated mass function to better than 10$\%$ over the range of -1.7 $\leq
\ln\sigma^{-1} \leq$ 0.5 while it  
appears to significantly overpredict haloes for $\ln\sigma^{-1} \geq$ 0.5.  
J01 also found an overprediction by the ST function
for $\ln\sigma^{-1} \geq 0.75$, which with their larger simulation
volumes, corresponded primarily to objects of
z$\leq$2 and of much higher mass.  Additionally, J01 found the mass function
to be invariant with redshift within their own results.  

R03 also 
considered the possibility of an empirical adjustment to the ST function. They inserted a crude multiplicative
factor to the ST function 
as follows, with $\delta_c$ $=$ 1.686 and FOF $ll=$0.2:
\begin{equation}
f(\sigma) = f(\sigma; {\rm S{\rm}T})\bigg[exp[-0.7/(\sigma
[\cosh(2\sigma)]^5)]\bigg],
\label{eq:reed}
\end{equation}
valid over the range of -1.7 $\leq
\ln\sigma^{-1} \leq$ 0.9.  The resulting function
is virtually identical to the ST function
for all $-\infty \leq  \ln\sigma\leq 0.4$. At higher values
of $\ln\sigma^{-1}$, this function declines relative to the ST
function, reflecting an underabundance of haloes that becomes greater
with increasing $\ln\sigma^{-1}$.  For -1.7 $\leq \ln\sigma^{-1} \leq$
0.5, Eq. (\ref{eq:reed}) matches R03 data to better than 10$\%$ for well-sampled
bins, while  for 0.5 $\leq \ln\sigma^{-1} \leq$ 0.9, where poisson
errors are larger, data is matched  to roughly 20$\%$. 
Note that in their notation, similarly to J01:
\begin{equation}\label{deff}
f(\sigma, z) \equiv \frac{M}{{\rho_b(z)}}{{\rm d}n(M, z)\over{\rm
d}\ln\sigma^{-1}}
\end{equation}
Using the relation $\nu=(\frac{\delta_{\rm co}}{\sigma})^2$ and Eq. (\ref{eq:universall}) one finds that:
\begin{equation}
f(\sigma,z)=2 \nu f(\nu)
\end{equation}

The evolution of the mass function can be calculated evaluating it at different redshifts. The mass function, Eq. (\ref{eq:universall}), depends on redshift because of the dependence of ${\rho_b(z)}$, $\nu$ or $\sigma$ from $z$. 
 
For example $\sigma(M,z) = \sigma(M,z=0)b(z)$, where ${\it b(z)}$ evolves as
$(1+z)^{-1}$ in an $\Omega_0 = 1$ universe, and more slowly in a
$\Lambda$CDM universe.  In the following I shall plot the mass function of the present paper and I compare it with R03 results.

\begin{figure}
\centerline{\hbox{
\psfig{file=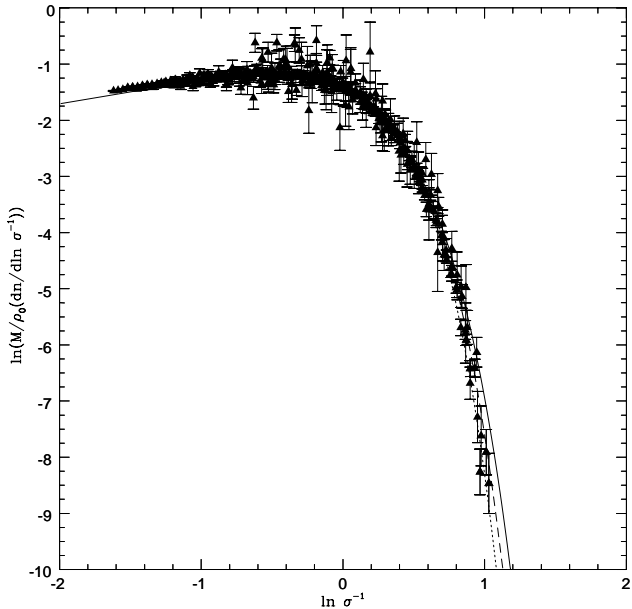,width=9cm}
}}
\caption[]{Mass function plotted in redshift independent form for all of
R03 outputs: redshifts used are 0, 1., 2., 3., 4., 5., 6.2,
7.8, 10., 12.1, 14.5. The solid line is ST prediction while the dashed and dotted line represent the result of the present paper and 
Eq. (\ref{eq:reed}), respectively.
}
\end{figure}

In Fig. 6, I plot the mass function for all of our outputs in the $f(\sigma)- \ln(\sigma^{-1})$ plane.  Large values of $\ln\sigma^{-1}$
correspond to rare haloes of high redshift and/or high  mass, while
small values of $\ln\sigma^{-1}$ describe haloes of low mass and
redshift  combinations.  
The solid line is the ST mass function while the dashed line the one obtained in the present paper and the dotted line represents 
Eq. (\ref{eq:reed}). 
The ST function fits the simulated mass function to better than 10$\%$ over the range of -1.7 $\leq
\ln\sigma^{-1} \leq$ 0.5 while it  
appears to significantly overpredict haloes for $\ln\sigma^{-1} \geq$ 0.5.  
The magnitude of the ST overprediction at
high values of ln$\sigma^{-1}$ is consistent with being a function
purely of ln$\sigma^{-1}$ rather than redshift, a natural consequence
of the fact that  the mass function is self similar in time (e.g
Efstathiou et al. 1988; Lacey \& Cole 1994; J01).
J01 also found an overprediction by the ST function
for $\ln\sigma^{-1} \geq 0.75$, which with their larger simulation
volumes, corresponded primarily to objects of
z$\leq$2 and of much higher mass.  Additionally, J01 found the mass function
to be invariant with redshift within their own results.  
The empirical adjustment to the ST mass function (Eq. (\ref{eq:reed})), dotted line, describes much better numerical simulations data: for -1.7 $\leq \ln\sigma^{-1} \leq$
0.5, Eq. (\ref{eq:reed}) matches R03 data to better than 10$\%$ for well-sampled
bins, while  for 0.5 $\leq \ln\sigma^{-1} \leq$ 0.9, where poisson
errors are larger, data is matched  to roughly 20$\%$. 
The ST and the mass function of the present paper differs more in the high mass region, where the mass function of the present paper is steeper than ST
and in better agreement with numerical simulations data than ST mass function. The better agreement between R03 simulations and the result of the present paper is 
connected to the improved barrier taking explicitly account the effects of several physical effects: namely dynamical friction, tidal torques and the cosmological constant.

\section{Conclusions}

In this paper, I have studied the role of non-radial motions, dynamical
friction and a non-zero cosmological constant on the collapse
of density peaks solving numerically the equations of motion of a shell of
baryonic matter falling into the central regions of a cluster of galaxies.
I have shown that non-radial motions and dynamical friction 
produce a delay in the collapse of
density peaks having low value of $\nu$ while the collapse of density peaks
having $\nu > 3$ is not influenced. 
Both effects produce a delay in collapse, but the effect of non-radial motions is slightly larger than that of dynamical friction.
A first consequence of this effect is a
reduction of the mass bound to collapsed perturbations and a raising of the
critical threshold, $\delta_{c}$, which now is larger than that of the
top-hat spherical model and depends on $\nu$. This means that shells of
matter of low density have to be subjected to a larger gravitational
potential, with respect to the homogeneous SM, in order to collapse.
The delay in the proto-structures collapse gives rise to a dynamical bias
similar to that described in CAD.
Finally, I used the Extended Press-Schechter (EPS) and ST1 approach to calculate the mass 
function starting from the threshold function. 
Then, I compared the numerical mass function given in R03 with the theoretical mass function obtained by means of the excursion set model and an improved version of the barrier shape obtained in the present paper, which implicitly takes account of dynamical friction, tidal interactions between clusters and a non-zero cosmological constant.
I showed that the barrier obtained in the present paper
gives rise to a better description of the mass function evolution with respect to other models (ST, ST1) and that the 
agreement is based on sound theoretical models and not on fitting to simulations like in R03, YNY, or Warren et al. (2005).

The main results of this last part of the paper can be summarized as follows: \\
1) the non-constant barrier of the present paper combined with the ST1 model gives a mass function evolution in better agreement with the N-body simulations of R03 
than other previous models (ST, ST1). \\
2) The mass function of the present paper gives a good fit to simulations results as the fit function proposed by R03, but differently from that it was obtained from a
sound theoretical background.\\
3)The excursion set model with a moving barrier is very versatile since it is very easy to introduce easily several physical effects in the 
calculation of the mass function and its evolution, just modifying the barrier.\\
4) The behavior of the mass function and its evolution at small masses is similar to that of ST, ST1, but at higher values of mass or redshift it is steeper than 
ST, ST1 in agreement with N-body simulations of R03.\\


The above considerations show that the excursion set approach that incorporates a non-spherical collapse which 
takes account of angular momentum acquisition, dynamical friction and a non-zero cosmological constant 
gives accurate predictions for a number of statistical quantities associated with the formation and clustering of dark matter haloes.
The improvement is probably connected also to the fact that incorporating the non-spherical collapse with increasing barrier in the excursion set approach results in a model in which fragmentation and mergers may occur, effects important in structure formation.
Moreover, the effect of a non-zero cosmological
constant adds to that 
of angular momentum and dynamical friction
%
slightly changing the evolution of the multiplicity function with respect to
open models with the same value of matter density parameter. 




\end{document}